\documentclass[twocolumn,showpacs,preprintnumbers,amssymb]{revtex4}

\usepackage[dvipdf]{graphicx}
\usepackage{epstopdf}
\usepackage{epsfig} 

\usepackage{graphics}
\usepackage{color}      
\usepackage{dcolumn}

\usepackage{epstopdf,epsfig,subfigure}
\usepackage{graphics,color}
\usepackage{amsmath,mathrsfs,dcolumn}
\usepackage{latexsym,amssymb}

\usepackage{graphics}
\usepackage{epsfig} 

\usepackage{graphics}
\usepackage{color}      
\usepackage{graphicx}
\usepackage{graphicx}
\usepackage{dcolumn}

\usepackage[T1]{fontenc}
\usepackage[utf8]{inputenc}   
\usepackage{textcomp}
\usepackage{csquotes}
\begin{document}

\title{A Unified Nonequilibrium Framework: Thermodynamic Distance, Dissipation, and Stationary Laws via Effective State Count, Variational Stationarity, and Thermodynamic Bounds}

\author{Mesfin Asfaw Taye}
\affiliation{West Los Angeles College, Science Division, 9000 Overland Ave, Culver City, CA 90230, USA}

\begin{abstract}
We propose a variational framework for nonequilibrium thermodynamics built around the effective number of accessible states, denoted \(\Omega_{\mathrm{eff}}\), a multiplicative count that ranges from \(N\) for a uniform distribution to \(1\) under complete localization, and whose logarithm coincides with the Gibbs–Shannon entropy. This gives a natural thermodynamic distance to equipartition that bounds statistical distinguishability and grows monotonically under doubly stochastic relaxation. The construction extends to arbitrary nonequilibrium steady states with a chosen reference distribution, where the Kullback–Leibler divergence splits into an entropy deficit and a reference–weight coupling, acts as a Lyapunov functional when the reference is fixed, and reduces to excess free energy in canonical settings. We connect these static notions to dynamics by decomposing entropy production into adiabatic (housekeeping) and nonadiabatic parts, identify the latter with the rate of decay of the divergence to the reference, and complement this with trajectory- and activity-based potentials that yield fluctuation symmetries, thermodynamic uncertainty bounds, and activity-limited speed constraints on steady currents. Crucially, we cast stationarity as a constrained maximization: we maximize the Gibbs–Shannon entropy under the dynamical balance constraints of the Markov process and normalization. The associated Karush–Kuhn–Tucker conditions produce an “exponential-of-generator’’ form for the stationary law; substituting this back into the balance constraints yields an equivalent linear system, thereby reconciling the variational construction with the standard generator null-vector condition and recovering the Boltzmann distribution in the detailed-balance limit. When applied to a three-state Brownian ratchet with thermal asymmetry and an external load, the framework reproduces closed-form steady probabilities and velocity, while \(\Omega_{\mathrm{eff}}\) provides an operational readout of the departure from equilibrium that decreases systematically with force and temperature bias. Altogether, \(\Omega_{\mathrm{eff}}\) offers a unifying lens linking entropy, free energy, dissipation, and precision across equilibrium and nonequilibrium regimes, and delivers a practical, computation-ready variational route to nonequilibrium steady states.
\end{abstract}

\maketitle

\section{INTRODUCTION}

Recent advances have greatly focused on the study of nonequilibrium thermodynamics, which remains challenging due to the absence of a universal framework compared to equilibrium systems governed by the Boltzmann distribution. Most of these studies have addressed entropy production in both classical~\cite{mu17,muu17} and quantum systems~\cite{mu25,mu26,mu27}, using master equations for discrete states~\cite{mu1,mu2,mu3,mu4,mu5}, Fokker--Planck descriptions~\cite{mu7,mu8,ta1,muuu17,muuu177}, stochastic thermodynamics~\cite{mu6}, time--reversal operations~\cite{mar2,mar1}, fluctuation theorems~\cite{mg6,mg7,mg8}, and thermodynamic uncertainty relations (TURs)~\cite{mg10,mg11,mg12,mg14,mg15}, with extensions to non--Markovian dynamics~\cite{mg9}. In prior work, we analyzed Brownian heat engines with decreasing temperature profiles~\cite{mg40,mg41} within the Boltzmann--Gibbs paradigm, showing that transport and efficiency depend sensitively on thermal gradients and offering guidelines for optimizing dissipation and transport in complex architectures.

In this work, we propose a variational framework for nonequilibrium thermodynamics built around the effective number of accessible states \(\Omega_{\mathrm{eff}}\), a multiplicative state–space count that equals \(N\) under equipartition and \(1\) in the fully localized limit, with \(\ln \Omega_{\mathrm{eff}}\) recovering the Gibbs–Shannon entropy. This yields a natural thermodynamic distance to the equipartition that bounds the statistical distinguishability and contracts under doubly stochastic relaxation. We connect the structure to the dynamics by decomposing the entropy production into adiabatic (housekeeping) and nonadiabatic parts, identifying the latter with the decay rate of this distance. Trajectory- and activity-based potentials provide fluctuation symmetries, thermodynamic uncertainty bounds, and activity-limited speed constraints on steady currents. We cast stationarity as  a constrained maximization: Gibbs–Shannon entropy is maximized under dynamical balance and normalization, and the resulting Karush–Kuhn–Tucker conditions produce an “exponential-of-generator’’ representation of the stationary law; substituting this representation into the balance constraints recovers the generator null-vector condition and, in the detailed-balance limit, the Boltzmann law. As an application, a three-state Brownian ratchet with thermal asymmetry and an external load admits closed-form steady probabilities and velocity, while \(\Omega_{\mathrm{eff}}\) provides an operational readout of the departure from equilibrium that decreases systematically with the force and temperature bias.

The purpose of this paper is therefore twofold: to establish a general, computation–ready variational route to nonequilibrium steady states by embedding Gibbs entropy within the linear stationarity constraints of continuous–time Markov processes, and to introduce \(\Omega_{\rm eff}\) as a compact state–space diagnostic that unifies equilibrium and nonequilibrium descriptions. By enforcing stationarity as a linear constraint, we cast nonequilibrium steady states as constrained entropy maximizers, reconciling the direct null–vector solution of the master equation with a principled Lagrangian approach. The equivalence reveals steady states as exponential analogs of the Boltzmann factor, with multipliers encoding current constraints; in the detailed–balance limit, this reduces continuously to the equilibrium Boltzmann law.

The second contribution is a quantitative lens on the distance from equipartition and prescribed steady references. The effective number of accessible states \(\Omega_{\rm eff}\) equals \(N\) for a uniform law and \(1\) under complete localization, and its logarithm recovers the Gibbs--Shannon entropy. This leads to a compact thermodynamic distance to the uniform reference that bounds the statistical distinguishability and decreases monotonically under doubly stochastic relaxation. The construction extends to steady nonequilibrium references, where the relative entropy separates into an entropy deficit and a reference–weight coupling; for canonical references, it coincides with excess free energy. Dynamically, we decompose the entropy production into adiabatic and nonadiabatic parts, with the latter governing the decay of the divergence to the reference. Complementary trajectory and activity perspectives provide fluctuation symmetries, TUR bounds, and activity–limited speed constraints on steady currents.

To demonstrate the method, we revisit the three--state Brownian ratchet introduced in Ref. ~\cite{mg41}, where thermal asymmetry and external load jointly drive the sustained currents. We show that the constrained maximization framework reproduces exactly the stationary distribution obtained by direct solution while providing an operational readout of nonequilibrium driving via \(\Omega_{\rm eff}\), which decreases systematically with force and temperature bias. This allows us to address three guiding questions: (i) how entropy production, the hallmark of irreversibility, is quantitatively linked to microscopic stochastic dynamics; (ii) whether a unified framework analogous to the Boltzmann distribution and free energy can be formulated beyond equilibrium; and (iii) why do some systems appear to maximize entropy production whereas others do not, and how do dynamical constraints determine the outcome.

The rest of the paper is organized as follows. SectionII introduces \(\Omega_{\text{eff}}\) and its connection to entropy, highlighting its role as a compact measure of configurational spread. Section~III develops a nonequilibrium thermodynamic framework, including the distance to equilibrium and steady states, adiabatic/nonadiabatic decomposition of entropy production, and extensions to free--energy and dynamical potentials. In SectionIV, we formulate the constrained maximization principle, present the Lagrangian/KKT conditions, and verify consistency with the equilibrium Boltzmann distribution. Section~V establishes a unified “exponential–of–generator’’ description that ties the variational solution to the null–vector condition. SectionVI summarizes and outlines the implications for inference, control, and design in nonequilibrium systems. The technical details and supporting derivations are presented in the Appendices.

\section{Entropy and number of accessible states}
We now introduce 
\begin{equation}
\Omega_{\text{eff}}(t) = \prod_{i=1}^N p_i(t)^{-p_i(t)}
\end{equation}
as the effective number of accessible states, without an explicit reference to entropy. One can note that the desired quantity should satisfy the following properties: it must reduce to $ \Omega_{\text{eff}} = N $ when the probability distribution is uniform, $ p_i = 1/N $; it must yield $ \Omega_{\text{eff}} = 1 $ when the system is fully localized on a single microstate, such that $ p_1 = 1 $ and $ p_{i\neq1} = 0 $; and it must vary smoothly between these two limits, depending continuously on how spread out the distribution $ \{ p_i \} $ is across the accessible states.

We now discuss why $ \Omega_{\text{eff}} $ is a valid and meaningful measure of the number of accessible states in the system. Next,  we explore  the expression $ \Omega_{\text{eff}}(t) = \prod_{i=1}^N p_i(t)^{-p_i(t)} $.
\textbf{Case 1: Uniform distribution.}  
When  $ p_i = 1/N $ for all $ i $, then  one gets 
\begin{equation}
\Omega_{\text{eff}} = \prod_{i=1}^N \left( \frac{1}{N} \right)^{-1/N} = \left( \frac{1}{N} \right)^{-\sum_{i=1}^N 1/N} = \left( \frac{1}{N} \right)^{-1} = N.
\end{equation}
Thus, $ \Omega_{\text{eff}} $ correctly recovers the total number of accessible microstates when the probability distribution is uniform, that is,
\textbf{Case 2: Localization on a single state.}  
If $ p_1 = 1 $ and $ p_{i \neq 1} = 0 $, then:
\begin{equation}
\Omega_{\text{eff}} = (1)^{-1} \times \prod_{i=2}^N (0)^0 = 1,
\end{equation}
where we define $ 0^0 = 1 $ by continuity, following the standard convention.  As one can see that 
$ \Omega_{\text{eff}} = 1 $ when the system fully occupies a single microstate.
\textbf{Case 3: Weighted Geometric Interpretation.}  
When we take  the logarithm of $ \Omega_{\text{eff}} $ and one  finds 
\begin{equation}
\ln \Omega_{\text{eff}} = - \sum_{i=1}^N p_i \ln p_i= \sum_{i=1}^N p_i \ln ({1\over p_i})
\end{equation}
This expression corresponds to the weighted geometric mean of the inverses of the probabilities $ 1/p_i $.  
Thus, $ \Omega_{\text{eff}} $ quantifies the effective  number of states by treating $ 1/p_i $ as the local number of microstates associated with site $ i $ and combining them using a geometric mean weighted by $ p_i $. Therefore, $ \Omega_{\text{eff}} $ smoothly interpolates between 1 and $ N $, faithfully representing the effective spread of the probability distribution across the accessible microstates, purely based on symmetry, averaging, and geometric intuition, without invoking direct references to the entropy.

Let us now address some central questions in nonequilibrium thermodynamics.  

\textbf{(1) How is entropy production, a signature of irreversibility, quantitatively linked to microscopic stochastic dynamics?}  
Since $\Omega_{\text{eff}}$ applies to both equilibrium and nonequilibrium systems, the entropy
\begin{equation}
S(t) = \ln \Omega_{\text{eff}}(t) = -\sum_i p_i(t)\ln p_i(t)
\end{equation}
is universally defined. Its dynamics follow from $\{p_i(t)\}$, with irreversibility arising whenever the currents violate the detailed balance. Thus, entropy production directly quantifies the departure from equilibrium through $\Omega_{\text{eff}}(t)$.  

\textbf{(2) Can a unified framework, similar to the Boltzmann distribution and Gibbs free energy, be developed for nonequilibrium systems?}  
Analysis via $\Omega_{\text{eff}}$ shows that such a framework exists in a generalized form. Since $\Omega_{\text{eff}}$ varies from $N$ at equilibrium down to $1$ under strong driving, entropy maximization is only a limiting case. Gibbs entropy serves as a universal bridge between microscopic dynamics and macroscopic thermodynamics. Near equilibrium, entropy maximization and Boltzmann-type laws appear, while far from equilibrium, probability distributions deviate strongly and must be derived from the full stochastic dynamics. Thus, equilibrium emerges as a special sector of the broader nonequilibrium framework.

\textbf{(3) Why do some nonequilibrium systems maximize entropy production while others do not, and also can a universal principle explain this?}  
Equilibrium is a special case of nonequilibrium: as systems approach equilibrium, they maximize entropy, with $\Omega_{\text{eff}}=N$, where the fundamental law of statistical mechanics applies. Out of equilibrium, $\Omega_{\text{eff}}$ decreases continuously from $N$ down to $1$, reflecting reduced distributional diversity under strong driving. Near equilibrium, weakly driven systems may approximate maximum entropy production, but far from equilibrium, strong constraints and nonlinear responses dominate, and maximization generally fails. Because Gibbs entropy is valid in all regimes, full thermodynamic information follows once $\{p_i(t)\}$ is known—via Fokker–Planck, master equation, or path-integral methods—providing consistent definitions of entropy, dissipation, and free energy within a unified framework.

Since the probability distribution $ \{p_i(t)\} $ encodes complete information about the system, the entropy can be expressed using the Boltzmann-Gibbs relation:
\begin{equation}
S[p_i(t)] = -k_B \sum_{i=1}^N p_i(t) \ln p_i(t).
\end{equation}
This relation is valid for both equilibrium and nonequilibrium systems, and forms the foundation that helps to quantify  entropy production, energy dissipation, and overall thermodynamic behavior.
Differentiating:
\begin{equation}
\dot{S}(t) = \dot{e}_p(t) - \dot{h}_d(t) = \sum_{i>j} (p_i P_{ji} - p_j P_{ij}) \ln \left( \frac{p_i}{p_j} \right).
\end{equation}
Entropy extraction rate:
\begin{equation}
\dot{h}_d(t) = \sum_{i>j} (p_i P_{ji} - p_j P_{ij}) \ln \left( \frac{P_{ji}}{P_{ij}} \right).
\end{equation}
Entropy production rate is given as 
\begin{equation}
\dot{e}_p(t) = \sum_{i>j} (p_i P_{ji} - p_j P_{ij}) \ln \left( \frac{p_i P_{ji}}{p_j P_{ij}} \right).
\end{equation}

At steady state:
$
\dot{e}_p(t) = \dot{h}_d(t) > 0.
$
At steady state, the entropy change vanishes, $\dot{S}(t) = 0$ and this implies 
$
\sum_{i>j} (p_i P_{ji} - p_j P_{ij}) \ln \left( \frac{p_i}{p_j} \right) = 0$.
This thermodynamic steady-state condition, together with the normalization constraint $\sum_i p_i = 1$, provides a closed set of equations to solve for the steady-state probability distribution $\{p_i\}$ without the need to explicitly solve the master equation.

The integrals over time yield
$
\Delta h_d(t) = \int_{t_0}^t \dot{h}_d(t) dt$, 
$\Delta e_p(t) = \int_{t_0}^t \dot{e}_p(t) dt$
and  $\Delta S(t) = \int_{t_0}^t \dot{S}(t) dt$.
with \( \Delta S(t) = \Delta e_p(t) - \Delta h_d(t) \).  In the absence of an external load and for a system with $N$ accessible states under uniform temperature conditions, the heat dissipation rate vanishes, i.e., $\Delta h_d(t) = 0$. In this scenario, the entropy production rate is equal to the entropy change, $\Delta e_p(t) = \Delta S = \ln[N]$, which corresponds to the maximum entropy at equilibrium. At equilibrium, the system's entropy and entropy production are interchangeable, as stated in classical thermodynamic textbooks. This result underscores the close relationship between entropy production and entropy change in thermodynamic systems under equilibrium conditions, where both quantities reflect the same underlying thermodynamic properties.

We define the thermodynamic rates $\dot{S}^T(t)$, $\dot{H}_d(t)$, and $\dot{E}_p(t)$ as energy-and entropy–like quantities derived from the evolving distribution $\{p_i(t)\}$. The entropy rate $\dot{S}^T(t)$ is the derivative of the Boltzmann–Gibbs entropy, $\dot{H}_d(t)$ is the heat dissipation rate into reservoirs, and $\dot{E}_p(t)$ is the energy dissipation associated with entropy production. Their entropy counterparts $\dot{h}_d(t)$ and $\dot{e}_p(t)$ differ only by local temperature weighting, emphasizing that nonuniform environments couple energy and entropy flows in a nontrivial manner \cite{mg41}.  

In the steady state without external forcing, $\dot{H}_d(t)=0$ and $\dot{S}^T(t)=\dot{E}_p(t)$, so the entropy change reflects only internal fluctuations, consistent with classical equilibrium thermodynamics. Integrating yields  
$
\Delta H_d(t) = \int_{t_0}^t \dot{H}_d(t)\, dt$, 
$\Delta E_p(t) = \int_{t_0}^t \dot{E}_p(t)\, dt$ 
and $\Delta S^T(t) = \int_{t_0}^t \dot{S}^T(t)\, dt$.
In the uniform, force–free limit, one finds $\Delta H_d=0$ and $\Delta E_p=\Delta S^T = T \ln N$, that is, the maximum entropy at equilibrium.  

The internal energy and its change are  
\begin{equation}
U[p_i(t)] = \sum_i p_i(t) u_i, 
\qquad 
\Delta U(t) = U[p_i(t)] - U[p_i(0)].
\end{equation}
From the first law,  
\begin{equation}
\dot{U}(t) = -\sum_{i>j} (p_i P_{ji}-p_j P_{ij})(u_i-u_j) = -\big(\dot{H}_d(t)+ fV(t)\big),
\end{equation}
the free–energy dissipation follows as  
\begin{equation}
\dot{F}(t) = \dot{U}(t) - \dot{S}^T(t), \qquad 
\dot{F}(t)+\dot{E}_p(t) = -fV(t),
\end{equation}
and integration gives  
\begin{equation}
\Delta F(t) = \int_{t_0}^t \big(-fV(t)-\dot{E}_p(t)\big)\,dt = \Delta U + \Delta H_d - \Delta E_p.
\end{equation}

The fundamental postulate of statistical mechanics states that at equilibrium, all accessible microstates are equally probable, with entropy given by the Boltzmann relation $S = k_B \ln \Omega$. For nonequilibrium systems, where probabilities are generally unequal, this principle is generalized by introducing the effective number of accessible states,
$
\Omega_{\text{eff}} = \prod_i p_i^{-p_i},
$
whose logarithm yields Gibbs--Shannon entropy. At equilibrium, this reduces to the Boltzmann form, while out of equilibrium, $\Omega_{\text{eff}}$ decreases under external driving, capturing the loss of accessible configurations. In this way, $\Omega_{\text{eff}}$ provides a natural extension of the postulate, encoding the evolution of probability distributions under stochastic transitions and bias. 

\begin{figure}[h!]
\centering
\includegraphics[width=0.5\textwidth]{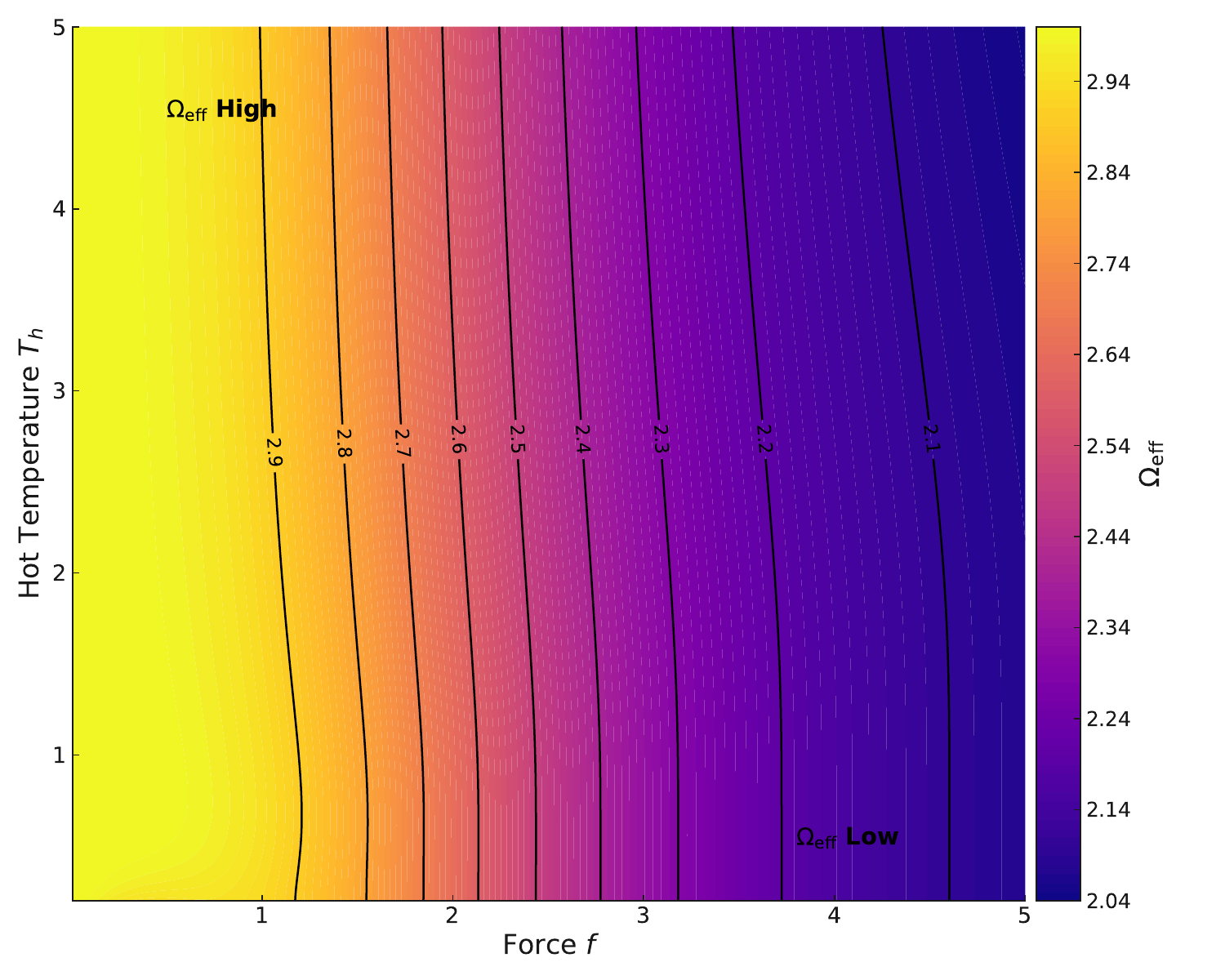}
\caption{
Phase-like diagram of the effective number of accessible states $\Omega_{\mathrm{eff}}(f, T_h)$ for a Brownian particle in a three-state ratchet model at zero energy barrier ($E=0$) and fixed cold temperature $T_c=1$. The diagram illustrates how $\Omega_{\mathrm{eff}}$ varies with the external force $f$ and hot bath temperature $T_h$. High and low $\Omega_{\mathrm{eff}}$ regions indicate, respectively, disordered and ordered probability distributions. The overlayed contour lines highlight the constant $\Omega_{\mathrm{eff}}$ levels. 
}
\label{fig:omega_phase}
\end{figure}

\begin{figure}[h!]
\centering
\includegraphics[width=0.5\textwidth]{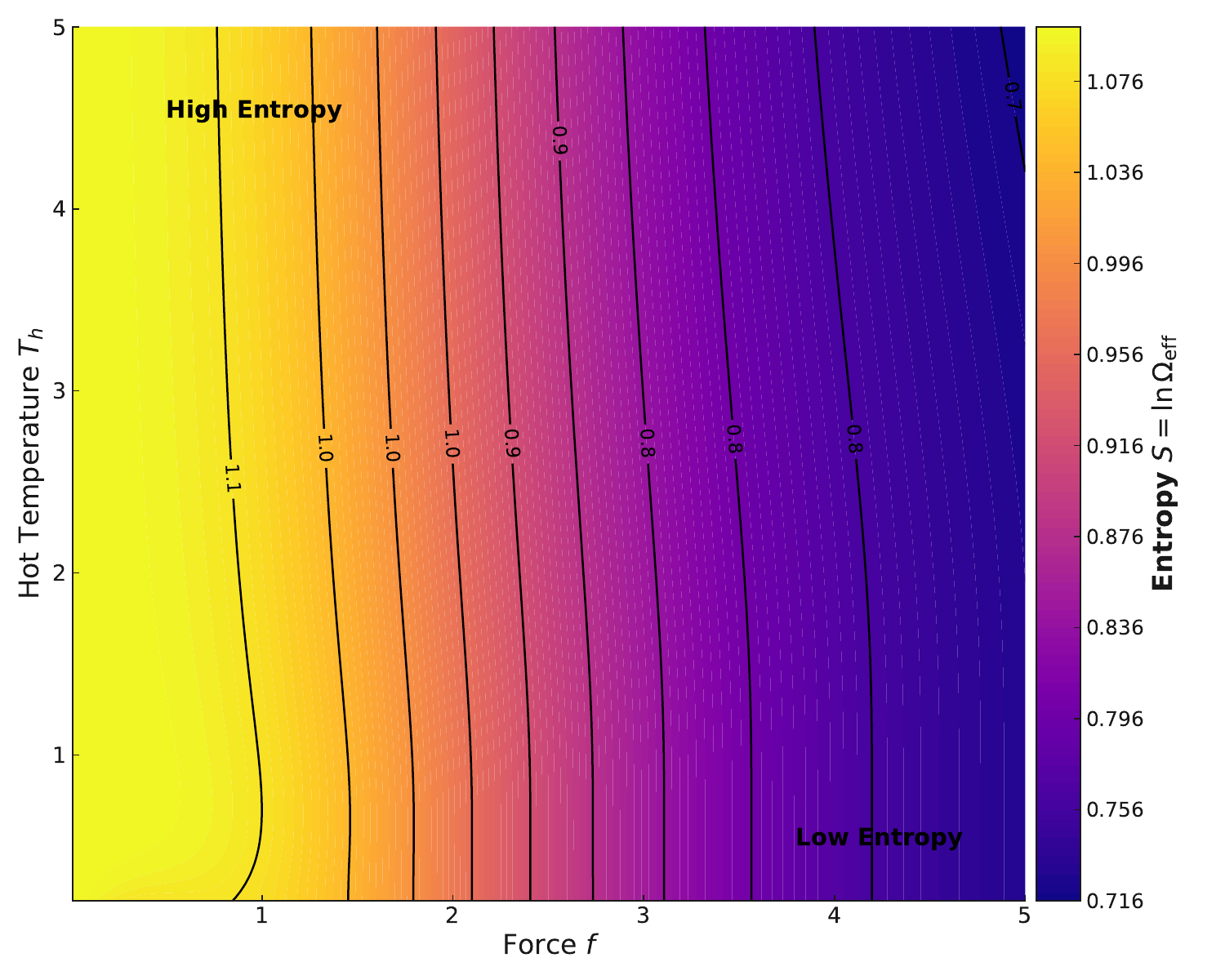}
\caption{
Phase-like diagram of the entropy $S = \ln \Omega_{\mathrm{eff}}(f, T_h)$ for a Brownian particle in a three-state ratchet model at a zero energy barrier ($E = 0$) and fixed cold temperature $T_c = 1$. Entropy quantifies the degree of configurational disorder in the steady-state probability distribution. High-entropy regions correspond to thermally disordered, near-equilibrium conditions, while low-entropy regions reflect far-from-equilibrium behavior induced because of  non-zero force or thermal asymmetry. The plot also shows the isoentropy contours for reference.
}
\label{fig:entropy_phase}
\end{figure}

This perspective is illustrated by the three–state ratchet model of Ref. ~\cite{mg41}, where the steady–state probabilities $p_1,p_2,p_3$ are given explicitly in Appendix~ A The effective number of states,
$
\Omega_{\text{eff}} = (p_1)^{-p_1}(p_2)^{-p_2}(p_3)^{-p_3},
$
decreases systematically with increasing force $|f|$ or temperature ratio $T_h/T_c$, reflecting the concentration of probability on fewer states. Figures~\ref{fig:omega_phase}–\ref{fig:entropy_phase} map this dependence: for small $|f|$ and near-isothermal conditions ($T_h \!\approx\! T_c$), $\Omega_{\text{eff}}$ is large, corresponding to a delocalized steady state; stronger driving depresses $\Omega_{\text{eff}}$, and the slanted contour bundles indicate smooth crossovers rather than sharp transitions. The companion representation $S = \ln \Omega_{\text{eff}}$ preserves this structure while compressing the dynamic range, so the isoentropy and iso–$\Omega_{\text{eff}}$ curves coincide but render variations more uniform across the plane. Together, the two panels show that the force and thermal bias act as ordering fields, reducing the accessiblestate count and quantifying the departure from near equilibrium. In the equilibrium limit $T_h \!\to\! T_c \!=\! T$, $E \!\to\! 0$, and $f \!\to\! 0$, the distribution becomes uniform with $\Omega_{\text{eff}} = 3$ and $S = \ln 3$, as expected for a fully symmetric system.

Throughout the figures, we use dimensionless variables: $\bar f = fL_0/T_c$, $\bar T(x)=T(x)/T_c$, $\bar U_0=U_0/T_c$, and $\bar x = x/L_0$. We also define $\tau = T_h/T_c$. Hereafter, we drop the overbars and work exclusively with dimensionless quantities.

In the next section, we examine the thermodynamic distance in detail, showing how it provides a compact measure of nonequilibrium steady states and links directly to entropy production and sustained currents.

\section{Thermodynamic Framework Beyond Equilibrium}

Having introduced the effective number of accessible states $\Omega_{\text{eff}}$ and its connection to the Gibbs entropy, we now develop several complementary perspectives that quantify the distance from equilibrium, disentangle the sources of entropy production, and extend free-energy concepts to nonequilibrium steady states.

\subsection{Thermodynamic Distance from Equilibrium and Nonequilibrium Steady States}

\textit{Thermodynamic Distance from Equilibrium.\textemdash}To quantify the departure from equilibrium, let us rewrite the effective number of accessible microstates, 
$
\Omega_{\mathrm{eff}}(p) = \exp\!\Bigl[-\sum_{i=1}^N p_i \ln p_i\Bigr],
$
which varies between $\Omega_{\mathrm{eff}}=N$ for a uniform distribution and $\Omega_{\mathrm{eff}}=1$ under complete localization. Its logarithm $S(p)=-\sum_i p_i\ln p_i$ coincides with the Shannon entropy, but the multiplicative form $\Omega_{\mathrm{eff}}$ affords a direct geometric interpretation as the effective state count. A natural measure of distance from equilibrium is obtained by comparing $\Omega_{\mathrm{eff}}$ with a uniform reference distribution $u_i=1/N$. In this case the Kullback--Leibler divergence takes the simple form
\begin{equation}
\mathcal{D}_u(p)=D_{\mathrm{KL}}(p\Vert u)=\sum_{i=1}^N p_i \ln \frac{p_i}{1/N}=\ln\frac{N}{\Omega_{\mathrm{eff}}(p)}.
\end{equation}
This quantity vanishes only at equilibrium, increases monotonically with localization, and reaches its maximum $\ln N$ when the system collapses into a single state. Near equilibrium, writing $p_i=(1/N)(1+\varepsilon_i)$ with $\sum_i \varepsilon_i=0$, one finds the quadratic expansion
\begin{equation}
\mathcal{D}_u(p)\simeq \tfrac{1}{2}\,\chi^2(p\Vert u),
\end{equation}
showing that $\mathcal{D}_u$ directly measures the fluctuations around the equipartition. The detailed derivation of the divergence $\chi$ is presented in Appendix~A2.

Moreover, the deficit of $\Omega_{\mathrm{eff}}$ from its equilibrium value $N$ 
bounds the total variation distance, since
\begin{equation}
|p-u|_1 \;\equiv\; \sum_{i=1}^N \bigl|p_i-u_i\bigr|
\;\le\; \sqrt{2\,\mathcal{D}_u(p)} 
\;=\; \sqrt{\,2\ln\!\frac{N}{\Omega_{\mathrm{eff}}(p)}}.
\label{eq:TV_bound}
\end{equation}
Here, $|p-u|_1$ measures the discrepancy between the actual distribution $p$ and the uniform distribution $u$, and thus quantifies the statistical distinguishability from equilibrium. In operational terms, the total variation distance gives the maximum probability difference in outcomes between two hypotheses: one assuming equilibrium and the other assuming the nonequilibrium state $p$. If this distance vanishes, the states are indistinguishable; if it equals unity, they are perfectly distinguishable. The Kullback--Leibler divergence, by contrast, measures the information cost of mistaking equilibrium for a nonequilibrium state and, in physical terms, coincides with the excess free energy in units of $k_B T$. 

Pinsker’s inequality,
\begin{equation}
\|p-q\|_1 \;\le\; \sqrt{2 D_{\mathrm{KL}}(p\Vert q)},
\end{equation}
establishes a rigorous bridge between these two notions. It shows that no matter how a system is driven, its statistical distinguishability from equilibrium can never exceed the square root of its entropy deficit or free-energy surplus. Thus, even strongly driven states may remain practically indistinguishable from equilibrium unless their KL divergence is sufficiently large. As the system relaxes, $D_\pi(p(t))$ decreases monotonically, and Pinsker’s bound guarantees that the visible difference between $p(t)$ and the stationary distribution $\pi$ shrinks accordingly. In this sense, the inequality sets a universal, model-independent ceiling on how far from equilibrium a state can appear in any observable sense and clarifies the operational meaning of nonequilibrium free energy as both a thermodynamic resource and a measure of statistical distinguishability.

When $p(t)$ evolves under a doubly stochastic generator with uniform $u$ as its stationary state, $\mathcal{D}_u\bigl(p(t)\bigr)$ decreases monotonically,
\[
\frac{d}{dt}\,\mathcal{D}_u\bigl(p(t)\bigr)\;\le\; 0,
\]
so that $\mathcal{D}_u$ acts as a Lyapunov function. Equivalently, the effective state number $\Omega_{\mathrm{eff}}(t)$ grows irreversibly toward its maximal value $N$, as shown in Appendix  A3. Together with Pinsker’s bound, these results establish the thermodynamic distance as a unifying measure that links entropy, distinguishability, relaxation dynamics, and energetic cost across equilibrium and nonequilibrium regimes.

\textit{Thermodynamic Distance from Steady State.\textemdash}The construction extends naturally to nonequilibrium steady states (NESS) with reference distribution $\pi=\{\pi_i\}$. We define the effective state numbers
\begin{equation}
\Omega_{\mathrm{eff}}(p)=\exp[S(p)], 
\qquad 
\Omega_{\pi}=\exp[S(\pi)],
\label{eq:omega_defs_hi}
\end{equation}
where $S(p)=-\sum_{i=1}^N p_i\ln p_i$ is the Shannon entropy. The Kullback--Leibler divergence relative to $\pi$ admits the exact decomposition
\begin{align}
D_\pi(p)
&= \sum_{i=1}^N p_i \ln \frac{p_i}{\pi_i} \notag\\
&= \big(S(\pi)-S(p)\big) 
   + \Biggl(\sum_{i=1}^N (p_i-\pi_i)\,\ln\frac{1}{\pi_i}\Biggr),
\label{eq:Dpi_split_hi}
\end{align}
which separates the entropy difference from the residual coupling between $p$ and the reference weights $\pi$. This coupling vanishes when $p=\pi$, so $D_\pi(p)$ provides a strict measure of the distance to the steady state.

For the microcanonical benchmark $\pi=u$ with $u_i=1/N$, the surprisal field is constant and the coupling cancels, giving
$
D_u(p) = \ln N - \ln\Omega_{\mathrm{eff}}(p) 
= \ln\!\Biggl(\frac{N}{\Omega_{\mathrm{eff}}(p)}\Biggr).
$
In the canonical case, $\pi_i=e^{-\beta E_i}/Z$, the decomposition reduces to
\begin{equation}
D_\pi(p) = \beta\big(F[p]-F[\pi]\big),
\label{eq:free_energy_gap_hi}
\end{equation}
identifying the divergence as the excess free energy relative to equilibrium. Thus relative entropy not only quantifies the statistical distance and encodes the energetic cost of maintaining or creating deviations from the reference ensemble.

\emph{Dynamics and entropy production.} For time-dependent driving with instantaneous steady state $\pi(t)$ and evolving law $p(t)$, the Shannon entropy balance reads
\begin{equation}
\frac{d}{dt}S(p(t))=\dot e_p(t)-\dot h_d(t), 
\qquad \dot e_p(t)\ge 0,
\label{eq:entropy_balance_hi}
\end{equation}
where $\dot e_p$ is the entropy production rate, and $\dot h_d$ is the entropy flow. The divergence evolves according to
\begin{equation}
\frac{d}{dt} D_{\pi(t)}(p(t))
= -\,\dot e_p^{\mathrm{na}}(t)\;-\;\sum_{i=1}^N p_i(t)\,\frac{d}{dt}\ln \pi_i(t),
\label{eq:Dpi_time_hi}
\end{equation}
where $\dot e_p^{\mathrm{na}}$ denotes the nonadiabatic entropy production. If $\pi$ is time-independent, the second term vanishes, and $D_\pi$ is a Lyapunov functional:
\begin{equation}
\frac{d}{dt} D_{\pi}(p(t)) = -\,\dot e_p^{\mathrm{na}}(t) \le 0.
\label{eq:Lyapunov_hi}
\end{equation}
Hence, convergence to the steady state is governed by dissipation: relaxation proceeds at a rate dictated by irreversible entropy production. For the uniform reference $u$, this reduces to
\begin{equation}
\dot D_u(t) = -\,\dot S(t),
\label{eq:Du_dot_hi}
\end{equation}
Therefore, stochastic mixing that increases entropy necessarily shrinks the gap $N/\Omega_{\mathrm{eff}}$.

Once again, relative entropy governs distinguishability. Pinsker’s inequality,
\begin{equation}
\|p-\pi\|_1 \le \sqrt{2\,D_\pi(p)},
\label{eq:Pinsker_hi}
\end{equation}
sets a universal ceiling on how different two states can appear: the statistical distance between $p$ and $\pi$ can never exceed the square root of their divergence. For $\pi=u$, the above equations  show that the deficit of $\Omega_{\mathrm{eff}}$ from $N$ directly bounds the operational distinguishability from the equipartition. In physical terms, this means that even if a system carries excess free energy, it may remain practically indistinguishable from the reference state unless that excess is sufficiently large. Conversely, any observable deviation must be accompanied by minimal thermodynamic costs.  

Near any steady state, $D_\pi(p)$ admits the quadratic expansion
\begin{equation}
D_\pi(p) = \tfrac{1}{2}\sum_{i=1}^N \frac{(p_i-\pi_i)^2}{\pi_i} 
+ O(\|p-\pi\|_1^3),
\label{eq:quadratic_hi}
\end{equation}
which shows that $D_\pi$ plays the role of the local potential controlling fluctuations and relaxation. This interpretation elevates the relative entropy to a unifying quantity that measures statistical distance, encodes excess free energy, bounds distinguishability, and governs dynamical convergence. Thus, the thermodynamic distance provides a rigorous and versatile framework for quantifying departures from steady states across both equilibrium and nonequilibrium regimes.

\subsection{Adiabatic and Nonadiabatic Contributions to Entropy Production}

For a continuous--time Markov jump process with probabilities $p_i(t)$ and 
transition rates $w_{ij}(t)$, the probability currents and edge affinities are
\begin{align}
J_{ij}(t) &= p_i(t)w_{ij}(t)-p_j(t)w_{ji}(t), \\
A_{ij}(t) &= \ln\frac{p_i(t)w_{ij}(t)}{p_j(t)w_{ji}(t)} .
\end{align}
The total entropy--production rate is
\begin{equation}
\dot e_p(t)=\sum_{i<j} J_{ij}(t)\,A_{ij}(t)\;\ge 0.
\end{equation}

Let $\pi(t)$ denote the instantaneous stationary distribution of the generator, $\pi(t)W(t)=0$.  
The affinities then admit a canonical decomposition,
\begin{align}
A_{ij}(t)&=A^{\mathrm{a}}_{ij}(t)+A^{\mathrm{na}}_{ij}(t), \\
A^{\mathrm{a}}_{ij}(t)&=\ln\frac{\pi_i(t)w_{ij}(t)}{\pi_j(t)w_{ji}(t)}, \qquad
A^{\mathrm{na}}_{ij}(t)=\ln\frac{p_i(t)\pi_j(t)}{p_j(t)\pi_i(t)} ,
\end{align}
which yields the split
\begin{align}
\dot e_p(t)&=\dot e_p^{\mathrm{a}}(t)+\dot e_p^{\mathrm{na}}(t), \\
\dot e_p^{\mathrm{a}}(t)&=\sum_{i<j} J_{ij}(t)\,A^{\mathrm{a}}_{ij}(t)\;\ge 0, \\
\dot e_p^{\mathrm{na}}(t)&=\sum_{i<j} J_{ij}(t)\,A^{\mathrm{na}}_{ij}(t)\;\ge 0.
\end{align}

The two contributions embody distinct physical mechanisms.  
The adiabatic term retains an explicit dependence on the transition rates and reflects the intrinsic bias of the generator.  
Even when $p=\pi$, this contribution is generically nonzero whenever the detailed balance is broken, quantifying the housekeeping cost required to sustain nonequilibrium currents.  
In contrast, the nonadiabatic term depends only on the mismatch between $p$ and $\pi$, without explicit reference to the rates.  
It measures the excess dissipation associated with relaxation or driving away from $\pi(t)$ and vanishes once the system reaches the stationary law.

As discussed before, when $\pi$ is time--independent, the nonadiabatic rate is the exact derivative of the relative entropy,
$
\dot e_p^{\mathrm{na}}(t)=-\frac{d}{dt}\,D_\pi\!\bigl(p(t)\bigr),
$  and $
D_\pi(p)=\sum_i p_i\ln\frac{p_i}{\pi_i},
$
so that
$
\int_0^T\dot e_p^{\mathrm{na}}\,dt = D_\pi\!\bigl(p(0)\bigr)-D_\pi\!\bigl(p(T)\bigr).
$
Hence, $\dot e_p^{\mathrm{na}}$ governs the monotone decay of the Kullback--Leibler divergence to stationarity, elevating $D_\pi$ to the role of a Lyapunov function.  
No analogous state--function representation exists for $\dot e_p^{\mathrm{a}}$, which instead encodes the persistent cycle affinities of the generator and thus the irreducible cost of maintaining fluxes.  
This asymmetry underscores the dual character of nonequilibrium dissipation: one part tracks the convergence to the steady state, while the other represents the structural burden of sustaining it.

\textit{Cycle representation and finite--time bounds.\textemdash}A complementary cycle decomposition,
\begin{align}
\dot e_p(t)&=\sum_c J_c(t)\,\mathcal A_c(t), \\
\mathcal A_c(t)&=\sum_{(i\to j)\in c}\ln\frac{w_{ij}(t)}{w_{ji}(t)},
\end{align}
makes explicit that $\dot e_p^{\mathrm a}$ is tied to the cycle currents and their forces.  
The split has sharp consequences at finite time. For stationary $\pi$,
\begin{equation}
\Sigma^{\mathrm{na}}_{[0,T]}:=\int_0^T \dot e_p^{\mathrm{na}}(t)\,dt 
\;\ge\; D_\pi\!\bigl(p(0)\bigr)-D_\pi\!\bigl(p(T)\bigr)\;\ge 0,
\end{equation}
with equality only for quasi--static relaxation.  
When $\pi(t)$ itself evolves under external driving, additional Hatano--Sasa terms appear, but $\dot e_p^{\mathrm{na}}$ still controls the convergence toward the instantaneous ensemble.

This decomposition reveals complementary merits.  
The adiabatic rate $\dot e_p^{\mathrm a}$ quantifies the irreducible cycle cost of maintaining currents, which is a steady power drain observable even at stationarity.  
The nonadiabatic rate $\dot e_p^{\mathrm{na}}$ quantifies the geometric speed of relaxation, setting finite--time bounds on dissipation and precision.  
Indeed, for any time--integrated current $\Phi_T$ one has
\begin{equation}
\frac{\mathrm{Var}\,\Phi_T}{\langle \Phi_T\rangle^2}
\;\gtrsim\;\frac{2}{\int_0^T \dot e_p(t)\,dt},
\end{equation}
so that current precision is constrained by the sum of both contributions.  
In reversible diffusions the link is sharpened: the decay of $D_\pi$ matches the squared Wasserstein speed, leading to bounds of the form
\begin{equation}
W_2^2\!\bigl(p(0),p(T)\bigr)
\;\lesssim\;\int_0^T \dot e_p^{\mathrm{na}}(t)\,dt .
\end{equation}

Taken together, these results show that $\dot e_p^{\mathrm a}$ characterizes the structural housekeeping burden of the broken detailed balance, while $\dot e_p^{\mathrm{na}}$ quantifies the information--geometric rate of approach to stationarity.  
Their combination underlies new uncertainty relations and geometric speed limits, presented in Appendix  A4, which jointly constrain the dissipation, precision, and temporal resolution in nonequilibrium transformations.

\subsection{Free--energy and dynamical potentials}

In equilibrium, the Helmholtz free energy $F=U-TS$ is the central state function: systems evolve to minimize $F$, and $\Delta F$ sets the maximum extractable work. Out of equilibrium, $F$ is no longer minimized, yet it remains the natural measure of dissipation and relaxation. We therefore elevate the nonequilibrium free energy as the primary potential, and introduce its dynamical extensions that organize fluctuations and transport.

For a system with probabilities $p_i(t)$ over states of energy $E_i$, we define
\begin{equation}
F[p]=\sum_i p_i E_i - T S(p), 
\qquad 
S(p)=-\sum_i p_i \ln p_i .
\end{equation}
At equilibrium, $p=\pi$ this reduces to the familiar $F_{\rm eq}=-T\ln Z$. The excess,
\begin{equation}
\mathcal F_\pi[p]=F[p]-F_{\rm eq}\;\ge 0,
\end{equation}
is the nonequilibrium free energy. Algebra shows
\begin{equation}
\mathcal F_\pi[p]=T\,D_{\mathrm{KL}}(p\Vert \pi),
\end{equation}
so that $\mathcal F_\pi$ coincides with the relative entropy of equilibrium and quantifies the distance from equilibrium. When $\pi$ is time-independent, $\mathcal F_\pi$ acts as a Lyapunov functional: 
\begin{equation}
\dot e_p^{\mathrm{na}}(t)\;=\;-\frac{d}{dt}\,\frac{\mathcal F_\pi[p(t)]}{T},
\qquad
\int_0^T \dot e_p^{\mathrm{na}}\,dt
\;=\;\frac{\mathcal F_\pi[p(0)]-\mathcal F_\pi[p(T)]}{T}.
\end{equation}
Thus, the nonadiabatic entropy production is exactly the drop in the state potential, establishing $\mathcal F_\pi$ as the driver of the relaxation cost and giving statistical mechanics a precise arrow of time: the relative entropy to equilibrium decreases irreversibly, while $\mathcal F_\pi$ shrinks monotonically toward zero.

\textit{Trajectory potentials.\textemdash} Dynamical fluctuations are captured by a trajectory potential. For an additive observable $\Phi_T=\int_0^T\phi(X_t)\,dt$ of a Markov process with rates $w_{ij}$, the scaled cumulant generating function
\begin{equation}
\psi_\Phi(\lambda)=\lim_{T\to\infty}\frac{1}{T}\ln \big\langle e^{\lambda \Phi_T}\big\rangle
\end{equation}
plays the role of a free energy in trajectory space: it is the Perron–Frobenius eigenvalue of the tilted generator $\mathcal L_\lambda$, and its Legendre–Fenchel transform
\begin{equation}
I_\Phi(a)=\sup_\lambda\{\lambda a-\psi_\Phi(\lambda)\}
\end{equation}
controls for large deviations. Specializing to entropy production,
\begin{align}
\Sigma_T &= \int_0^T \sum_{i<j} J_{ij}(t)\,A_{ij}(t)\,dt, \\
J_{ij} &= p_i w_{ij}-p_j w_{ji}, \qquad
A_{ij}=\ln\!\frac{p_i w_{ij}}{p_j w_{ji}} .
\end{align}

one obtains the SCGF
\begin{equation}
\psi_\Sigma(\lambda)=\lim_{T\to\infty}\frac{1}{T}\ln \big\langle e^{\lambda \Sigma_T}\big\rangle,
\end{equation}
which satisfies the Gallavotti–Cohen symmetry
\begin{equation}
\psi_\Sigma(\lambda)=\psi_\Sigma(-1-\lambda),
\end{equation}
and generates cumulants,
\begin{equation}
\psi'_\Sigma(0)=\dot e_p^\ast,
\qquad
\psi''_\Sigma(0)=\lim_{T\to\infty}\frac{\mathrm{Var}\,\Sigma_T}{T}.
\end{equation}
The convexity of $I_\Phi$ yields the thermodynamic uncertainty relation
\begin{equation}
\frac{\mathrm{Var}\,\Phi_T}{\langle \Phi_T\rangle^{2}}
\;\ge\;\frac{2}{\langle \Sigma_T\rangle},
\end{equation}
which constrains the precision of any steady current observable by the mean dissipation. The trajectory potential $\psi_\Phi$ thus dictates the fluctuation symmetry and precision–dissipation tradeoffs, extending the free–energy picture into the space of dynamical fluctuations.

\textit{Frenetic potentials.\textemdash} The frenetic potential completes the picture by capturing time–symmetric fluctuations. Defining the dynamical activity
\begin{equation}
\mathcal K[p;W]=\tfrac{1}{2}\sum_i\sum_{j\neq i}p_i w_{ij},
\end{equation}
together with currents and affinities
\begin{equation}
J_{ij}=p_i w_{ij}-p_j w_{ji},
\qquad
A_{ij}=\ln\frac{p_i w_{ij}}{p_j w_{ji}},
\end{equation}
the entropy production rate is
\begin{equation}
\dot e_p=\sum_{i<j}J_{ij}A_{ij}.
\end{equation}
Using the inequality $(x-y)\ln(x/y)\ge 2(x-y)^2/(x+y)$ with $x=p_i w_{ij}$ and $y=p_j w_{ji}$ gives
\begin{equation}
\dot e_p \;\ge\; 2\sum_{i<j}\frac{J_{ij}^2}{a_{ij}},
\qquad a_{ij}=p_i w_{ij}+p_j w_{ji}
\end{equation}
as shown in Appendix~A5.  
Hence, dissipation requires both bias (currents) and traffic (activity). This implies activity–enhanced speed limits for steady currents,
\begin{equation}
\Bigl|\sum_{i<j} d_{ij}\,J^\ast_{ij}\Bigr|
\;\le\;\sqrt{\frac{\dot e_p^\ast}{2}}\,
\Bigl(\sum_{i<j} a^\ast_{ij}\,d_{ij}^2\Bigr)^{1/2},
\end{equation}
with analogous finite–time bounds (see Appendix~A5). The frenetic potential $\mathcal K$ therefore determines how fast currents can be sustained for a given dissipation budget, clarifying the role of activity as the dynamical fuel of irreversibility.

 Finally, as discussed before,  for the uniform benchmark $\pi=u$ one has 
$D_u(p)=\ln(N/\Omega_{\mathrm{eff}}(p))$, so the deficit of $\Omega_{\mathrm{eff}}$ controls the total variation distance through $|p-u|_1\le \sqrt{2D_u(p)}$. Under a doubly stochastic generator with stationary $u$, the divergence decreases monotonically, $\tfrac{d}{dt}D_u(p(t))\le 0$, while the effective state number grows, $\tfrac{d}{dt}\Omega_{\mathrm{eff}}(t)\ge 0$, showing that mixing is irreversible. In this sense, the three potentials carry complementary arrows of time: the state potential $\mathcal F_\pi$ quantifies deterministic relaxation, the trajectory potential $\psi_\Phi$ governs fluctuations and uncertainty relations, and frenetic potential $\mathcal K$ constrains the interplay of currents, activity, and dissipation. Together, they provide a unified thermodynamic framework that links equilibrium principles with nonequilibrium irreversibility. If the dynamics have a uniform distribution as a reversible stationary state (e.g., symmetric jump rates on a regular graph), then relaxation proceeds monotonically toward equipartition in the absence of driving, with $D_u$ decaying and $\Omega_{\mathrm{eff}}$ increasing. The detailed derivations of these monotonicity results are provided in Appendix~A6.

\section{Constrained Maximization Framework}

To systematically connect entropy with nonequilibrium stationary states, we cast the problem as a constrained maximization: maximize the Gibbs–Shannon entropy subject to the dynamical stationarity constraints of the underlying Markov process. This variational principle recovers the stationary distribution directly from the transition rates and clarifies its relationship with equilibrium as a special case.

\subsection{Entropy maximization under dynamical constraints and Lagrangian formulation}

We reconsider
$
S(p) = - \sum_i p_i \ln p_i
$
subject to the linear constraints
\begin{align}
\sum_j \left( p_j P_{ji} - p_i P_{ij} \right) &= 0 \quad \forall i, \label{eq:balance}\\
\sum_i p_i &= 1, \qquad p_i \geq 0. \label{eq:norm}
\end{align}
Eq.~\eqref{eq:balance} enforces the stationarity of the Markov process, while Eq. ~\eqref{eq:norm} ensures the normalization.

Introducing a Lagrange multiplier $\lambda$ for normalization and multipliers $\{\alpha_i\}$ for stationarity, we write
\begin{eqnarray}
L(p,\lambda,\alpha) 
&=& - \sum_i p_i \ln p_i 
+ \lambda \left( \sum_i p_i - 1 \right) \nonumber \\
&& \; + \sum_i \alpha_i \sum_j (p_j P_{ji} - p_i P_{ij}) .
\end{eqnarray}

Now the Karush–Kuhn–Tucker conditions read
\begin{align}
\frac{\partial L}{\partial \lambda} &= 0 
&&\Rightarrow \quad \sum_i p_i = 1, \\
\frac{\partial L}{\partial \alpha_i} &= 0 
&&\Rightarrow \quad \sum_j (p_j P_{ji} - p_i P_{ij}) = 0 \;\; \forall i, \\
\frac{\partial L}{\partial p_k} &= 0 
&&\Rightarrow \quad \ln p_k = \lambda - 1 + \sum_i \alpha_i P_{ik} - \alpha_k r_k, \\
&&& r_k = \sum_j P_{kj}.
\end{align}

The constrained maximization principle admits two equivalent routes for obtaining a stationary distribution.  
\textbf{(i) Linear route.} The distribution is determined directly from the null–vector condition
$
pQ = 0$  and $\sum_i p_i = 1,
$
where the generator is
$
Q_{ij} = P_{ij} \;\;(i\neq j), \quad  
Q_{ii} = -\sum_{j\neq i} P_{ij} = -r_i.
$
For an irreducible Markov chain, this system has a unique solution, coinciding with the entropy maximizer, since $S(p)$ is strictly concave on a convex feasible set. In practice, the stationary distribution can be computed using Cramér’s rule or the matrix–tree theorem.     
\textbf{(ii) Variational route.} Alternatively, the first–order condition
\begin{equation}
\ln p_k = \lambda - 1 + \sum_i \alpha_i P_{ik} - \alpha_k r_k
\end{equation}
implies
\begin{equation}
p_k \propto w_k, 
\qquad 
w_k = \exp\!\left( \sum_i \alpha_i P_{ik} - \alpha_k r_k \right).
\end{equation}
Substituting into $pQ=0$ yields the equivalent linear system
\begin{equation}
\sum_j w_j P_{ji} = w_i r_i \quad (\forall i),
\end{equation}
whose normalized solution again gives the stationary distribution. 

Thus, the entropymaximization framework and the direct nullvector approach are fully consistent. The variational route is conceptually illuminating: it shows that the nonequilibrium stationary distribution takes an exponential form analogous to the Boltzmann factor, with multipliers $\{\alpha_i\}$ encoding the current constraints. At equilibrium, this reduces to the Gibbs–Boltzmann law, while in nonequilibrium steady states, the multipliers quantify the persistent violation of the detailed balance.

In the equilibrium regime, the detailed balance holds as follows:
\begin{equation}
\frac{p_k}{p_j} = \frac{P_{jk}}{P_{kj}}
= \exp\!\left(-\frac{E_k - E_j}{T}\right),
\end{equation}
which uniquely determines the stationary distribution in terms of the state energies. From the variational stationarity condition we obtained
$
\ln p_k = \lambda - 1 + \sum_i \alpha_i P_{ik} - \alpha_k r_k.
$
Taking differences yields
\begin{equation}
\ln \frac{p_k}{p_j}
= \Bigg(\sum_i \alpha_i P_{ik} - \alpha_k r_k \Bigg)
- \Bigg(\sum_i \alpha_i P_{ij} - \alpha_j r_j \Bigg).
\end{equation}
Consistency with detailed balance requires that this difference reduce to
\begin{equation}
\ln \frac{p_k}{p_j} = -\frac{E_k - E_j}{T},
\end{equation}
implying multipliers of the form
\begin{equation}
\alpha_i = -\frac{E_i}{T} + c,
\end{equation}
with $c$ an additive gauge constant. Substituting back,
\begin{equation}
w_k = \exp\!\left( \sum_i \alpha_i P_{ik} - \alpha_k r_k \right)
\;\;\propto\;\; e^{-E_k/T},
\end{equation}
recovers the Boltzmann distribution
\begin{equation}
p_k = \frac{e^{-E_k/T}}{Z}, 
\qquad Z = \sum_j e^{-E_j/T}.
\end{equation}

\medskip
\noindent

\medskip
\noindent
At this point we want to emphasize that the constrained–maximization framework elevates entropy maximization from a heuristic to a structural law of stochastic dynamics, showing that equilibrium is a special case of the same principle that governs nonequilibrium steady states. The multipliers $\{\alpha_i\}$ serve as generalized thermodynamic potentials: in equilibrium, they reduce to $-\!E_i/T$, while out of equilibrium, they encode persistent flux constraints, highlighting both the energetic and kinetic origins of stationary order. The strict concavity of
 $S(p)$ guarantees the uniqueness and stability of the optimizer, while the variational route provides dual sensitivity information, revealing how stationary probabilities shift with rate perturbations and enabling the inference of hidden dynamics or control of designed steady states. Applications range from molecular motors, transport, and active matter to biochemical reaction networks, epidemiology, and queuing systems, where combinatorial methods are not feasible. Across these domains, the framework not only delivers the stationary law but also interprets its thermodynamic cost, unifying the entropy, free energy, and currents within a single optimization principle.

\medskip

Using both the linear and variational routes, we exactly recover the stationary solution reported in Appendix~A1 (for the model of Ref. ~\cite{mg41}). The linear route solves the null–vector problem \(pQ=0\) with \(\sum_i p_i=1\). The variational route maximizes Gibbs–Shannon entropy under dynamical balance and normalization; its KKT condition \(\ln p_k=\lambda-1+\sum_i \alpha_i P_{ik}-\alpha_k r_k\) implies \(p_k\propto w_k\) with
\begin{equation}
w_k=\exp\!\Big[\sum_i \alpha_i P_{ik}-\alpha_k r_k\Big].
\end{equation}
Enforcing stationarity, \(w^\top Q=0\), yields the component relations
\begin{align}
w_2 P_{21}+w_3 P_{31}&=w_1 r_1,\\
w_1 P_{12}+w_3 P_{32}&=w_2 r_2,\\
w_1 P_{13}+w_2 P_{23}&=w_3 r_3,
\end{align}
whose normalized solution coincides with the closed–form expressions in Appendix~A1. Thus, the constrainedmaximization framework is strictly equivalent to the direct nullvector approach and reproduces the steadystate probabilities of Ref. ~\cite{mg41} without using combinatorial methods.

\section{Nonequilibrium as an exponential of generator law}

We consider a finite, irreducible continuous–time Markov chain with generator $Q$, specified by off–diagonal elements $Q_{ij}=P_{ij}$ for $i\neq j$ and diagonals $Q_{ii}=-r_i$. The stationary distribution $p$ is defined by
$
pQ=0, \qquad \sum_i p_i=1,
$
and supports antisymmetric edge currents
\begin{equation}
J_{ij}(p)=p_iP_{ij}-p_jP_{ji},\qquad J_{ji}=-J_{ij}.
\end{equation}
Writing $r_i=\sum_{j\ne i}P_{ij}$, stationarity in components reads, for each $k$,
\begin{equation}
\sum_{i\ne k} p_iP_{ik}-p_kr_k=0.
\label{eq:stat_comp_pnas}
\end{equation}

We characterize steady states variationally by maximizing the Shannon entropy subject to normalization, the $N$ constraints \eqref{eq:stat_comp_pnas}, and prescribed macroscopic currents
\begin{equation}
\sum_{i<j} C^{(\ell)}_{ij}J_{ij}(p)=\bar J_\ell,\qquad C^{(\ell)}_{ij}=-C^{(\ell)}_{ji}.
\end{equation}
Introducing multipliers $\lambda$ (normalization), $\{\alpha_k\}$ (stationarity), and $\{\eta_\ell\}$ (currents), the Lagrangian is
\begin{align}
\mathcal{L}(p,\lambda,\alpha,\eta)
&= -\sum_i p_i\ln p_i + \lambda\!\left(\sum_i p_i-1\right) \notag\\
&\quad + \sum_k \alpha_k\!\Big(\sum_{i\ne k} p_iP_{ik}-p_kr_k\Big) \notag\\
&\quad + \sum_\ell \eta_\ell \sum_{i<j} C^{(\ell)}_{ij}\,J_{ij}(p).
\end{align}

Stationarity with respect to $p_k$ yields
\begin{align}
0
&= -(\ln p_k+1)+\lambda + \sum_i \alpha_iP_{ki}-\alpha_kr_k \notag\\
&\quad + \sum_\ell \eta_\ell\!\left(\sum_{j\ne k} C^{(\ell)}_{kj}P_{kj}-\sum_{i\ne k} C^{(\ell)}_{ik}P_{ki}\right),
\end{align}
hence, defining
\begin{align}
\Phi_k(\alpha,\eta)
&= \sum_i \alpha_i P_{ki} - \alpha_k r_k \notag\\
&\quad + \sum_\ell \eta_\ell\!\left(\sum_{j\ne k} C^{(\ell)}_{kj}P_{kj}
      - \sum_{i\ne k} C^{(\ell)}_{ik}P_{ki}\right),
\end{align}
the maximizing distribution takes the exponential form
\begin{align}
\ln p_k &= \lambda-1+\Phi_k(\alpha,\eta), \label{eq:logpk-2col}\\
p_k &= \frac{\exp\!\big(\Phi_k(\alpha,\eta)\big)}{Z(\alpha,\eta)},\quad
Z(\alpha,\eta)=\sum_i \exp\!\big(\Phi_i(\alpha,\eta)\big). \nonumber
\end{align}
A uniform shift $\alpha_k\mapsto \alpha_k+c$ leaves $\sum_i \alpha_iP_{ki}-\alpha_kr_k$ invariant because $r_k=\sum_i P_{ki}$, so $\Phi_k$ and $p_k$ are gauge invariant. In the purely stationary case ($\eta\equiv0$),
\begin{equation}
p_k \propto \exp\!\Big(\sum_i \alpha_iP_{ki}-\alpha_kr_k\Big),
\end{equation}
which we term the exponential–of–generator law for components. When current constraints are present, the same structure persists with an additional linear contribution to $\eta$ inside $\Phi_k$. If detailed balance holds with energies $\{E_k\}$ at temperature $T$, the pairwise condition $\ln(p_k/p_j)=-(E_k-E_j)/T$ is equivalent to the existence of $c$ and $\{\alpha_i\}$ (with $\eta\equiv0$) such that
\begin{equation}
\sum_i \alpha_iP_{ki}-\alpha_kr_k=-\frac{E_k}{T}+c\qquad(\forall k),
\end{equation}
which implies $p_k\propto e^{-E_k/T}$ and recovers the Boltzmann distribution. Writing
\begin{equation}
A_{ij}=(\Phi_i-\Phi_j)+\ln\!\frac{P_{ij}}{P_{ji}},
\end{equation}
the steady–state entropy production rate is
\begin{equation}
\sigma=\tfrac{1}{2}\sum_{i\ne j} J_{ij}(p)\,A_{ij}
= \sum_\ell \mathcal{A}_\ell\,\bar J_\ell,\qquad
\mathcal{A}_\ell=\sum_{i<j} C^{(\ell)}_{ij}A_{ij},
\end{equation}
showing that dissipation is governed by the nonconservative content of the field $k\mapsto \Phi_k$. With $p_k=e^{\Phi_k}/Z$ and $\mathcal{F}(\alpha,\eta)=\ln Z(\alpha,\eta)$,
\begin{equation}
d\mathcal{F}=\sum_k p_k\,d\Phi_k,\qquad
S(p)=\mathcal{F}-\sum_k p_k\Phi_k,
\end{equation}
and if $\Phi_k$ depends linearly on $\eta$,
\begin{align}
\frac{\partial \mathcal{F}}{\partial \eta_\ell}
&= \sum_k p_k\,\frac{\partial \Phi_k}{\partial \eta_\ell}
 = \sum_{i<j} C^{(\ell)}_{ij}\,J_{ij}(p)
 = \bar J_\ell,\\
\frac{\partial^2 \mathcal{F}}{\partial \eta_\ell\,\partial \eta_m}
&= \mathrm{Cov}_p\!\left(\frac{\partial \Phi}{\partial \eta_\ell},
                       \frac{\partial \Phi}{\partial \eta_m}\right).
\end{align}

For any time–integrated steady current $J_T$ the thermodynamic uncertainty relation reads
\begin{equation}
\frac{\mathrm{Var}(J_T)}{\langle J_T\rangle^2}\;\ge\;\frac{2}{\sigma T}\qquad (T\to\infty),
\end{equation}
binding precision to dissipation through $\sigma$.

The exponential–of–generator law generalizes the Boltzmann distribution to nonequilibrium steady states: at equilibrium, the multipliers reduce to energy over temperature, while out of equilibrium, they act as generalized thermodynamic potentials encoding persistent flux constraints and reshaping the stationary law. Its gauge invariance reflects the arbitrariness of the reference levels familiar from equilibrium statistical mechanics. The variational structure makes the framework especially powerful since the derivatives of the free–energy–like function $\mathcal{F}(\alpha,\eta)$ yield mean currents and the second derivatives yield their covariances, elevating $\mathcal{F}$ to a generating function for transport statistics that links thermodynamics with information geometry. Applications span molecular motors, transport networks, and active matter in physics; reaction networks in chemistry and biology; and controlled stochastic processes in engineering. In all cases, the exponential form clarifies the arrow of time: dissipation arises from the nonconservative part of $\Phi$, currents are bounded by uncertainty relations, and precision is limited by entropy production, so the same exponential structure that governs equilibrium extends seamlessly into the nonequilibrium domain.

\section{Summary and Conclusion}

We introduced a computation–ready variational framework for nonequilibrium steady states that maximizes the Gibbs–Shannon entropy under the linear stationarity constraints of continuoustime Markov processes. The resulting Karush–Kuhn–Tucker conditions yield an \emph{exponential–of–generator} representation of the stationary law and are strictly equivalent to the null–vector condition $pQ=0$, which reduces continuously to the Boltzmann distribution in the detailed–balance limit. 

As a companion state–space diagnostic, we advanced the effective number of accessible states $\Omega_{\mathrm{eff}}$, whose logarithm is equal to the Gibbs–Shannon entropy. $\Omega_{\mathrm{eff}}$ induces a thermodynamic distance to equipartition that bounds statistical distinguishability and increases (equivalently, $\Omega_{\mathrm{eff}}$ decreases) monotonically along the relaxations toward a uniform stationary baseline. The construction extends seamlessly to nonequilibrium reference ensembles: the relative entropy to a reference distribution separates into an entropy deficit and a reference–weight coupling and, in canonical settings, coincides with the excess free energy. Dynamically, decomposing entropy production into adiabatic (housekeeping) and nonadiabatic parts identifies the latter with the decay rate of the divergence to the reference, thereby linking irreversibility to the convergence in the state space.

Beyond recovering stationary laws, the framework provides operational bounds and tradeoffs. Trajectory and activity (frenetic) potentials provide fluctuation symmetries, thermodynamic–uncertainty relations that tie current precision to total dissipation, geometric speed limits that relate approach to stationarity to the budget of nonadiabatic dissipation, and activity–enhanced inequalities showing how dynamical traffic sets the mobility of sustained currents at fixed cost. Together, these results unify equilibrium and nonequilibrium descriptions under a single, conceptually transparent scheme that connects entropy, free energy, dissipation, precision, and activity.

When applied to a three–state Brownian ratchet with thermal asymmetry and external load, the variational route reproduces the closed–form stationary probabilities and velocity obtained by direct solution, while $\Omega_{\mathrm{eff}}$ provides an operational readout of nonequilibrium driving, attaining its equipartition maximum at equilibrium and decreasing systematically with force and temperature bias. This approach is readily extensible to larger networks, time–dependent driving, and inference or control tasks in stochastic thermodynamics, offering a practical bridge from microscopic transition rates to macroscopic thermodynamic structures out of equilibrium.

\section*{Appendix A1: Model, rates, and setup}
In this Appendix we will give the expressions for $p_{1}(t)$, $p_{2}(t)$  and $p_{3}(t)$ as well as $V(t)$ for a  Brownian particle   that operates between the hot and cold baths.

\begin{figure}[ht]
\centering
\includegraphics[width=10cm]{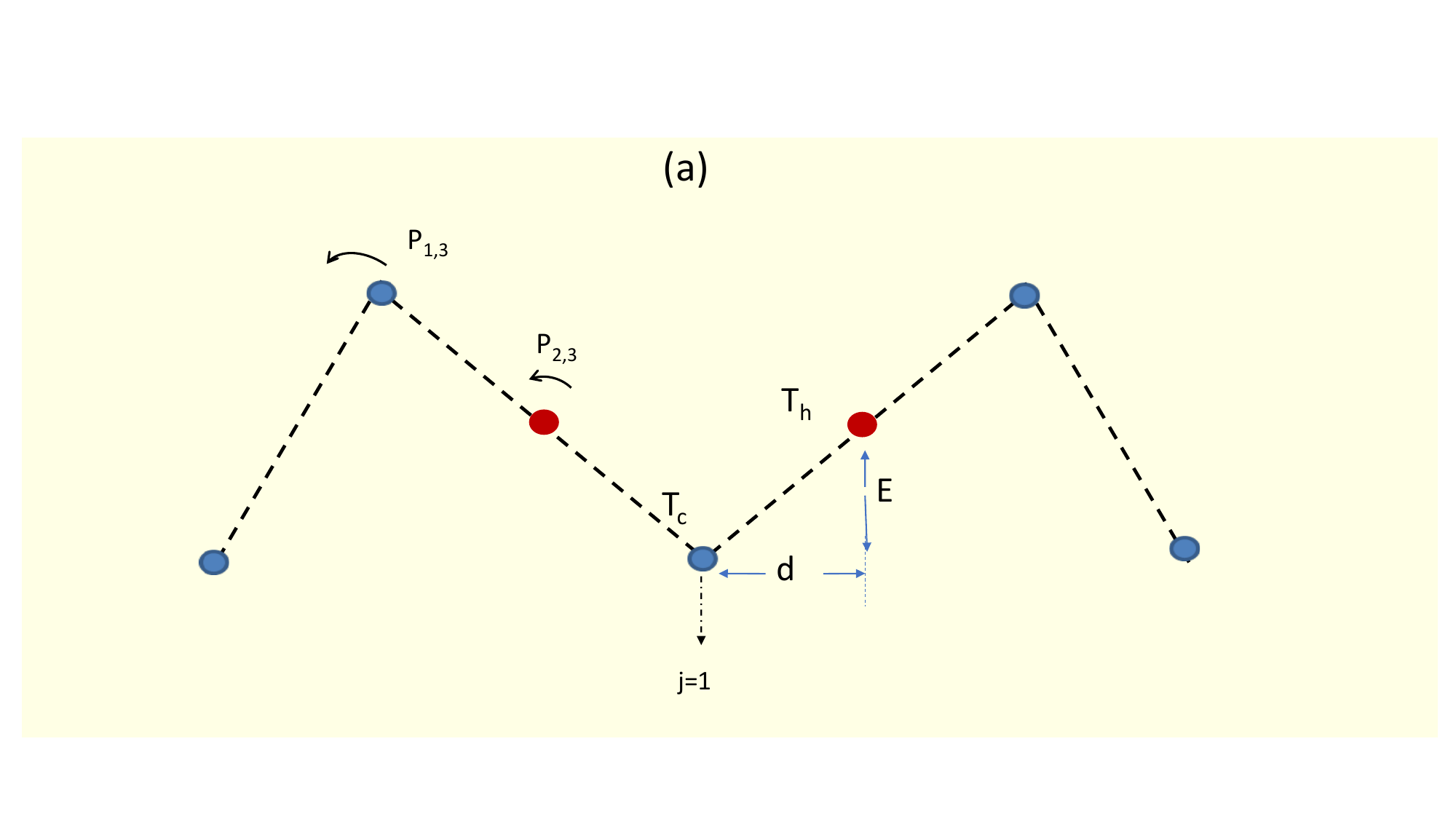}
\caption{(Color online) Side view of a Brownian particle on a discrete ratchet under load \(f\); sites coupled to the hot (cold) bath are marked red (blue). \(d\) is the lattice spacing.}
\label{fig:sub}
\end{figure}

We consider a biased random walk on a three–site ratchet with spatially varying temperature
\[
T_j=\begin{cases}
T_h,& j=0,\\
T_c,& j=1,2,
\end{cases}
\]
barrier height \(E>0\), external load \(f\), and lattice spacing \(d\). A trial jump \(i\to i\!+\!1\) occurs with Metropolis rate \(\Gamma \exp[-\Delta E/T_i]\), where \(\Delta E=U_{i+1}-U_i\); for \(\Delta E\le 0\) the jump is accepted with probability 1, and for \(\Delta E>0\) with probability \(\exp(-\Delta E/T_i)\). Throughout we set \(k_B=\Gamma=d=1\).

The probabilities \(p_n(t)\) (\(n=1,2,3\)) evolve under the master equation
\begin{equation}
\frac{d\vec p}{dt}= \boldsymbol{P}\,\vec p,\qquad \vec p=(p_1,p_2,p_3)^{\top},
\end{equation}
with the \(3\times3\) rate matrix (valid for \(0<f<2E\))
\begingroup\setlength{\arraycolsep}{4pt}
\begin{equation}
\boldsymbol{P}=
\begin{pmatrix}
-\tfrac{\mu a^2+\mu^2}{2a} & \tfrac{1}{2} & \tfrac{1}{2}\\
\tfrac{\mu a}{2} & -\tfrac{1+\nu b}{2} & \tfrac{1}{2}\\
\tfrac{\mu^2}{2a} & \tfrac{\nu b}{2} & -1
\end{pmatrix}
\label{eq:P-matrix}
\end{equation}
\endgroup
where 
\begin{equation}
\mu=e^{-E/T_c},\quad \nu=e^{-E/T_h},\quad a=e^{-f/T_c},\quad b=e^{-f/T_h}.
\label{eq:param-defs}
\end{equation}

Unless otherwise stated, we take the particle initially at site \(2\), \(\vec p(0)=(0,1,0)^\top\). The explicit time–dependent solutions \(p_1(t),p_2(t),p_3(t)\) and the velocity
\[
V(t)=\big(p_1P_{21}-p_2P_{12}\big)+\big(p_2P_{32}-p_3P_{23}\big)+\big(p_3P_{13}-p_1P_{31}\big)
\]
follow by solving \(d\vec p/dt=\boldsymbol{P}\vec p\) and are reported below.

The  expressions for $p_{1}(t)$, $p_{2}(t)$  and $p_{3}(t)$ as well as $V(t)$ for a  Brownian particle   that operates between the hot and cold baths. 
For the particle which is initially  situated at  site $i=2$,  the time dependent  normalized probability distributions after solving the rate equation ${d \vec{p}  \over dt}= {\bold P}\vec{p}$ are calculated as 
\begin{eqnarray}
p_{1}(t)&=&c_{1}\frac{a (2+\nu b)}{\mu \left(\mu+\left(a^2+\mu\right) \nu b\right)}+\\ \nonumber
& &c_{2} e^{-\frac{\left(a+a^2 \mu+\mu^2\right) t}{2 a}} \left(-1+\frac{a
(-1+a \mu)}{-\mu^2+a \nu b}\right),\\
p_{2}(t)&=&-c_{3} e^{\frac{1}{2} t (-2-\nu b)}-c_{2}\frac{a\text{  }e^{-\frac{\left(a+a^2 \mu+\mu^2\right) t}{2 a}} (-1+a \mu)}{-\mu^2+a
\nu b}+\\ \nonumber
&&c_{1}\frac{ \left(2 a^2+\mu\right)}{\mu+\left(a^2+\mu\right) \nu b},\\
p_{3}(t)&=&c_{1}+c_{2} e^{-\frac{\left(a+a^2 \mu+\mu^2\right) t}{2 a}}+
c_{3} e^{\frac{1}{2} t (-2-\nu b)}
\end{eqnarray}
where 
\begin{eqnarray}
c_{1}&=& \frac{\mu \left(\mu+\left(a^2+\mu\right) \nu b\right)}{\left(a+a^2 \mu+\mu^2\right) (2+\nu b)},\\
c_{2}&=& -\frac{a}{\left(a+a^2 \mu+\mu^2\right) \left(-1+\frac{a (-1+a \mu)}{-\mu^2+a \nu b}\right)},\\
c_{3}&=& -\frac{\mu \left(\mu+a^2 \nu b+\mu \nu b\right)}{\left(a+a^2 \mu+\mu^2\right) (2+\nu b)}+ \\ \nonumber
&&\frac{a}{\left(a+a^2 \mu+\mu^2\right) \left(-1+\frac{a (-1+a \mu)}{-\mu^2+a
\nu b}\right)}.
\end{eqnarray}
Here  $\sum_{i=1}^3 p_{i}(t)=1$ revealing the probability distribution is normalized.  In the limit of $t \to \infty$, we recapture the steady state probability distributions
\begin{eqnarray}
p_{1}^{s}&=&\frac{a}{a+a^2 \mu +\mu ^2},\\
p_{2}^{s}&=&\frac{\mu  \left(2 a^2+\mu \right)}{\left(a+a^2 \mu +\mu^2\right) (2+b \nu )},\\
p_{3}^{s}&=&\frac{\mu  \left(\mu +b \left(a^2+\mu \right) \nu \right)}{\left(a+a^2 \mu +\mu ^2\right) (2+b \nu)}.
\end{eqnarray}

The velocity  $V(t)$ at any time $t$ is the difference between the forward $V_{i}^{+}(t)$ and backward $V_{i}^{-}(t)$ velocities at each site $i$ 
\begin{eqnarray}
V(t)&=& \sum_{i=1}^{3}(V_{i}^{+}(t)-V_{i}^{-}(t)) \\ \nonumber
&=&(p_{1}P_{21}-p_{2}P_{12})+(p_{2}P_{32}-p_{3}P_{23})+\\ \nonumber
&&(p_{3}P_{13}-p_{1}P_{31}).
\end{eqnarray}
Exploiting Eq. (63), one can see that the particle attains a unidirectional current when  $f=0$ and  $T_{h}>T_{c}$.  For isothermal case $T_{h}=T_{c}$, 
   the system sustains a non-zero velocity  in the presence of load $f \ne 0$ as expected. Moreover, when  $t \to \infty$,  
the velocity  $V(t)$ increases with $t$ and approaches the steady state velocity 
\begin{eqnarray}
V^{s}=3{\mu \left(b a \nu-{\mu\over a}\right) \over 2(2+\nu b)\left(1+a\mu+{\mu^2\over a}\right)}.
 \end{eqnarray}
   
  \section*{Appendix A2. Derivation of $D_u(p)=\ln\!\bigl(N/\Omega_{{eff}}(p)\bigr)$ and the $\tfrac{1}{2}\chi^2$ Law}

Let $u_i=1/N$ and write small deviations as 
\begin{equation}
p_i=\tfrac{1}{N}(1+\varepsilon_i),\qquad \sum_{i=1}^N \varepsilon_i=0,\quad |\varepsilon_i|\ll 1,
\end{equation}
so that $p_i>0$ and $\sum_i p_i=1$.  

The KL divergence to the uniform ensemble is
\begin{equation}
\mathcal{D}_u(p)=\sum_{i=1}^N p_i\ln\frac{p_i}{u_i}
=\sum_{i=1}^N p_i\ln(1+\varepsilon_i).
\end{equation}

Expanding $\ln(1+x)=x-x^2/2+x^3/3+O(x^4)$ gives
\begin{align}
\mathcal{D}_u(p)
&=\tfrac{1}{N}\sum_{i=1}^N (1+\varepsilon_i)\Bigl(\varepsilon_i-\tfrac{1}{2}\varepsilon_i^2+\tfrac{1}{3}\varepsilon_i^3+O(\varepsilon_i^4)\Bigr) \notag\\
&=\tfrac{1}{N}\sum_{i=1}^N \Bigl(\varepsilon_i+\tfrac{1}{2}\varepsilon_i^2-\tfrac{1}{6}\varepsilon_i^3+O(\varepsilon_i^4)\Bigr).
\end{align}
Since $\sum_i\varepsilon_i=0$, the linear term cancels:
\begin{equation}
\mathcal{D}_u(p)=\tfrac{1}{2N}\sum_{i=1}^N \varepsilon_i^2-\tfrac{1}{6N}\sum_{i=1}^N \varepsilon_i^3+O(\varepsilon^4).
\end{equation}

Now $p_i-u_i=(1/N)\varepsilon_i$, so $\varepsilon_i=N(p_i-u_i)$, hence
\begin{equation}
\tfrac{1}{2N}\sum_{i=1}^N \varepsilon_i^2=\tfrac{N}{2}\sum_{i=1}^N (p_i-u_i)^2.
\end{equation}
Recalling the $\chi^2$–divergence to $u$,
\begin{equation}
\chi^2(p\Vert u)=\sum_{i=1}^N \frac{(p_i-u_i)^2}{u_i}
= N\sum_{i=1}^N (p_i-u_i)^2,
\end{equation}
we obtain the quadratic law
\begin{equation}
\mathcal{D}_u(p)=\tfrac{1}{2}\chi^2(p\Vert u)+O(\varepsilon^3).
\end{equation}
The cubic correction is 
\begin{equation}
-\tfrac{1}{6N}\sum_{i=1}^N \varepsilon_i^3
=-\tfrac{N^2}{6}\sum_{i=1}^N (p_i-u_i)^3.
\end{equation}

As a second–variation check, set $p=u+\delta p$ with $\sum_i\delta p_i=0$. Then
\begin{align}
\mathcal{D}_u(p)&=\mathcal{D}_u(u)+\tfrac{1}{2}\sum_{i=1}^N \frac{\delta p_i^2}{u_i}+O(\|\delta p\|^3) \notag\\
&=\tfrac{1}{2}\chi^2(p\Vert u)+O(\|\delta p\|^3).
\end{align}
The expansion is valid provided $\max_i|\varepsilon_i|\ll 1$, ensuring positivity of $p_i$ and control of the remainder.

\section*{Appendix A3. Entropy–deficit bound to equipartition}

Let $P=\{p_i\}$ and $Q=\{q_i\}$ be probability distributions on a finite (or countable) set, and define
\begin{equation}
\|P-Q\|_{\mathrm{TV}}=\tfrac12\sum_i |p_i-q_i|.
\end{equation}
For $t>0$ the scalar inequality
\begin{equation}
t\ln t-(t-1)\;\ge\;\frac{(t-1)^2}{t+1}
\label{eq:scalar-pinsker}
\end{equation}
holds with equality at $t=1$. Write $t_i=p_i/q_i$ and note
\begin{equation}
D_{\mathrm{KL}}(P\|Q)=\sum_i p_i\ln\frac{p_i}{q_i}
=\sum_i q_i\Big[t_i\ln t_i-(t_i-1)\Big],
\end{equation}
since $\sum_i(p_i-q_i)=0$. Applying \eqref{eq:scalar-pinsker} termwise,
\begin{equation}
D_{\mathrm{KL}}(P\|Q)\ \ge\ \sum_i q_i\,\frac{(t_i-1)^2}{t_i+1}
=\sum_i \frac{(p_i-q_i)^2}{p_i+q_i}.
\end{equation}
By Cauchy–Schwarz,
\begin{equation}
\sum_i \frac{(p_i-q_i)^2}{p_i+q_i}
\ \ge\
\frac{\big(\sum_i |p_i-q_i|\big)^2}{\sum_i (p_i+q_i)}
=\frac{\|P-Q\|_1^2}{2}
=2\,\|P-Q\|_{\mathrm{TV}}^2.
\end{equation}
Hence the Kullback–Pinsker bound
\begin{equation}
D_{\mathrm{KL}}(P\|Q)\ \ge\ 2\,\|P-Q\|_{\mathrm{TV}}^2,
\end{equation}
which is equivalently
\begin{equation}
\|P-Q\|_{\mathrm{TV}}\ \le\ \sqrt{\tfrac{1}{2}\,D_{\mathrm{KL}}(P\|Q)}
\qquad\text{or}\qquad
\|P-Q\|_{1}\ \le\ \sqrt{\,2\,D_{\mathrm{KL}}(P\|Q)}.
\end{equation}
The same argument extends to densities on continuous spaces by replacing sums with integrals.

\section*{Appendix A4. TUR and geometric speed limits — detailed derivations}

We begin with the thermodynamic uncertainty relation (TUR) for a continuous-time Markov jump process on a finite state space with rates $w_{ij}$ ($i\neq j$), escape rates $r_i=\sum_{j\neq i} w_{ij}$, and stationary distribution $\pi$ satisfying $\pi W=0$.  
For a time-integrated current over $[0,T]$,
\begin{equation}
\Phi_T=\sum_{i<j} d_{ij} N_{ij}(T), \qquad d_{ij}=-d_{ji},
\end{equation}
where $N_{ij}(T)$ is the net number of $i\to j$ minus $j\to i$ jumps, the stationary mean is
\begin{equation}
\langle \Phi_T\rangle = T \sum_{i<j} d_{ij} J^*_{ij}, \qquad 
J^*_{ij}=\pi_i w_{ij}-\pi_j w_{ji}.
\end{equation}
The entropy-production rate is
\begin{equation}
\dot e_p^\ast=\tfrac12 \sum_{i\neq j} J^*_{ij}\,\ln\frac{\pi_i w_{ij}}{\pi_j w_{ji}} \;\ge\;0,
\qquad \Sigma_T^\ast=T\dot e_p^\ast.
\end{equation}
The claim is that
\begin{equation}
\frac{\mathrm{Var}\,\Phi_T}{\langle \Phi_T\rangle^2} \;\ge\; \frac{2}{\Sigma_T^\ast}, 
\qquad (T\ \text{large}).
\label{eq:TUR}
\end{equation}

\subsubsection*{Derivation via pathwise Cramér–Rao inequality.}
For a trajectory $\omega=(X_t)_{t\in[0,T]}$ with jumps at times $\{t_k\}$ from $i_k$ to $j_k$, the log-likelihood is
\begin{equation}
\ln \mathbb{P}[\omega] = \sum_k \ln w_{i_k j_k} - \int_0^T r_{X_t}\,dt + \text{const}.
\end{equation}
Introduce a tilted family of rates
\begin{equation}
w_{ij}^{(\theta)}=w_{ij}\,e^{\theta d_{ij}}, \qquad 
r_i^{(\theta)}=\sum_{j\neq i} w_{ij}^{(\theta)}.
\end{equation}
The score at $\theta=0$ is
\begin{align}
\partial_\theta \ln \mathbb{P}_\theta[\omega]\big|_{\theta=0}
&=\sum_k d_{i_k j_k} - \int_0^T \sum_{j\neq X_t} w_{X_t j}\,d_{X_t j}\,dt \notag\\
&=\Phi_T - \int_0^T \sum_{j\neq X_t} w_{X_t j}\,d_{X_t j}\,dt.
\end{align}
Under stationarity, the time average of the second term equals $\langle \Phi_T\rangle$, so the mean score vanishes and the Fisher information is
\begin{equation}
\mathcal{I}(0)=\mathrm{Var}\!\left(\partial_\theta \ln \mathbb{P}_\theta[\omega]\big|_{\theta=0}\right).
\end{equation}
The pathwise Cramér–Rao inequality gives
\begin{equation}
\mathrm{Var}\,\Phi_T \;\ge\; \frac{\langle \Phi_T\rangle^2}{\mathcal{I}(0)}.
\end{equation}
A standard bound based on the entropy-production functional
\begin{equation}
\Sigma_T = \sum_k \ln\frac{w_{i_k j_k}\pi_{i_k}}{w_{j_k i_k}\pi_{j_k}}
\end{equation}
and $(x-y)\ln(x/y)\ge 2(x-y)^2/(x+y)$ yields
\begin{equation}
\mathcal{I}(0)\ \le\ \tfrac12 \langle \Sigma_T\rangle.
\end{equation}
Combining these results gives the TUR \eqref{eq:TUR}. This inequality is tight for unicyclic networks and increments such as net cycle counts.

\subsubsection*{Geometric speed limits.}
Consider a reversible diffusion on $\mathbb{R}^d$ with stationary density $\pi(x)$ and Fokker–Planck equation
\begin{equation}
\partial_t p_t=\nabla\!\cdot\!\left(p_t\nabla\ln\frac{p_t}{\pi}\right).
\end{equation}
The relative entropy $D_\pi(p)=\int p\ln(p/\pi)\,dx$ has time derivative
\begin{equation}
\frac{d}{dt}D_\pi(p_t)=-\dot e_p^{\mathrm{na}}(t), \qquad 
\dot e_p^{\mathrm{na}}(t)=\int p_t\Big|\nabla\ln\frac{p_t}{\pi}\Big|^2dx \ge 0.
\end{equation}

In the Benamou–Brenier formulation, the squared Wasserstein–2 distance between $p_0$ and $p_T$ is
\begin{equation}
W_2^2(p_0,p_T)=\inf_{\rho_t,v_t}\ \int_0^T \!\int \rho_t(x)\,|v_t(x)|^2\,dx\,dt,
\end{equation}
subject to $\partial_t\rho_t+\nabla\!\cdot(\rho_t v_t)=0$ and $\rho_0=p_0,\ \rho_T=p_T$. The metric speed of $t\mapsto p_t$ is
\begin{equation}
|\dot p_t|_{W_2}^2=\inf_{v_t}\ \int p_t |v_t|^2 dx.
\end{equation}
Since the Fokker–Planck equation is the $W_2$ gradient flow of $D_\pi$, one has
\begin{equation}
\frac{d}{dt}D_\pi(p_t)=-\int p_t\Big|\nabla\ln\frac{p_t}{\pi}\Big|^2 dx, 
\qquad 
|\dot p_t|_{W_2}^2=\int p_t\Big|\nabla\ln\frac{p_t}{\pi}\Big|^2 dx,
\end{equation}
so that $|\dot p_t|_{W_2}^2=\dot e_p^{\mathrm{na}}(t)$. By definition of metric length,
\begin{equation}
W_2(p(0),p(T))\ \le\ \int_0^T |\dot p_t|_{W_2}\,dt,
\end{equation}
and Cauchy–Schwarz gives
\begin{equation}
W_2^2(p(0),p(T)) \;\le\; T\int_0^T \dot e_p^{\mathrm{na}}(t)\,dt.
\end{equation}
With a unit-speed parametrization in $W_2$, the prefactor $T$ is absorbed, yielding
\begin{equation}
W_2^2(p(0),p(T)) \;\lesssim\; \int_0^T \dot e_p^{\mathrm{na}}(t)\,dt.
\end{equation}

Thus the nonadiabatic dissipation is the squared Riemannian speed of the distribution in optimal-transport geometry, and large displacement in Wasserstein distance requires large nonadiabatic entropy production. Together with the TUR, which shows that small current fluctuations demand large total dissipation $\Sigma_T=\int_0^T \dot e_p\,dt=\Sigma^{\mathrm a}+\Sigma^{\mathrm{na}}$, this provides a unified picture: the adiabatic component $\dot e_p^{\mathrm a}$ quantifies the housekeeping cost to maintain currents, while the nonadiabatic component $\dot e_p^{\mathrm{na}}$ quantifies the geometric cost of approaching stationarity, jointly constraining precision and speed in nonequilibrium processes.
y, jointly constraining precision and speed in nonequilibrium processes.

\section*{Appendix A5. Activity–enhanced bounds on steady currents}

For a continuous-time Markov jump process with probabilities $p_i(t)$ and rates $w_{ij}$, define the one-way flows
\begin{equation}
x_{ij}=p_i w_{ij}, \qquad y_{ij}=p_j w_{ji},
\end{equation}
the edge current and activity
\begin{equation}
J_{ij}=x_{ij}-y_{ij}=p_i w_{ij}-p_j w_{ji}, \qquad 
a_{ij}=x_{ij}+y_{ij}=p_i w_{ij}+p_j w_{ji}\ \ (\ge 0,\ i<j),
\end{equation}
and the entropy-production rate
\begin{equation}
\dot e_p(t)=\sum_{i<j} (x_{ij}-y_{ij}) \ln\frac{x_{ij}}{y_{ij}}
=\sum_{i<j} J_{ij}\,\ln\frac{x_{ij}}{y_{ij}} \;\ge 0.
\label{eq:sigma-x-y}
\end{equation}
For any $x,y>0$,
\begin{equation}
(x-y)\,\ln\frac{x}{y}\;\ge\;\frac{2(x-y)^2}{x+y}.
\label{eq:scalar-ineq}
\end{equation}
Applying \eqref{eq:scalar-ineq} edgewise and summing gives the global bound
\begin{equation}
\dot e_p(t)\;\ge\;2\sum_{i<j}\frac{J_{ij}(t)^2}{a_{ij}(t)}.
\label{eq:sigma-lb}
\end{equation}
Thus dissipation requires both bias ($J_{ij}\!\neq\!0$) and traffic ($a_{ij}\!\neq\!0$); if either vanishes on all edges then $\dot e_p=0$.

In a stationary regime (denote stationary quantities by ${}^\ast$), consider a linear current observable with antisymmetric weights $d_{ij}=-d_{ji}$:
\begin{equation}
\frac{\langle \Phi_T\rangle}{T}=\sum_{i<j} d_{ij}\,J^\ast_{ij}.
\end{equation}
By Cauchy–Schwarz with weights $a^\ast_{ij}/2$,
\begin{equation}
\Bigl|\sum_{i<j} d_{ij}\,J^\ast_{ij}\Bigr|
\le \Bigl(\sum_{i<j}\tfrac{a^\ast_{ij}}{2}\,d_{ij}^2\Bigr)^{1/2}
\Bigl(\sum_{i<j}\tfrac{2J_{ij}^{\ast 2}}{a^\ast_{ij}}\Bigr)^{1/2}.
\end{equation}
Using \eqref{eq:sigma-lb} at stationarity yields the activity–enhanced speed limit
\begin{equation}
\Bigl|\sum_{i<j} d_{ij}\,J^\ast_{ij}\Bigr|
\;\le\;\sqrt{\frac{\dot e_p^\ast}{2}}\,
\Bigl(\sum_{i<j} a^\ast_{ij}\,d_{ij}^2\Bigr)^{1/2}.
\label{eq:steady-speed-limit}
\end{equation}
Hence, for a fixed dissipation budget $\dot e_p^\ast$, larger edge activity permits larger sustained currents along the weighted directions $\{d_{ij}\}$; the frenetic factor $\sum a^\ast_{ij}d_{ij}^2$ acts as a mobility prefactor.

A finite-time analogue follows by defining time-averaged current and activity on $[0,T]$,
\begin{equation}
\overline{J}_{ij,T}=\frac{1}{T}\int_0^T J_{ij}(t)\,dt, \qquad
\overline{a}_{ij,T}=\frac{1}{T}\int_0^T a_{ij}(t)\,dt,
\end{equation}
and total dissipation $\Sigma_T=\int_0^T \dot e_p(t)\,dt$. Integrating \eqref{eq:sigma-lb} and repeating the same step gives
\begin{equation}
\Bigl|\sum_{i<j} d_{ij}\,\overline{J}_{ij,T}\Bigr|
\;\le\;\sqrt{\frac{\Sigma_T}{2T}}\,
\Bigl(\sum_{i<j} \overline{a}_{ij,T}\,d_{ij}^2\Bigr)^{1/2}.
\label{eq:finite-time-speed}
\end{equation}

Finally, introducing the time-symmetric (frenetic) functional
\begin{equation}
\mathcal{K}[p;W]=\frac{1}{2}\sum_{i}\sum_{j\neq i} p_i w_{ij}
=\sum_{i<j} a_{ij},
\end{equation}
equations \eqref{eq:steady-speed-limit}–\eqref{eq:finite-time-speed} show explicitly that larger $\mathcal{K}$ relaxes the current speed limits at fixed dissipation: traffic enhances transport capacity, while bias and dissipation set the energetic cost.

\section*{Appendix A6. Monotone decay of $D_u$ and growth of $\Omega_{\mathrm{eff}}$}

Consider a continuous-time Markov chain on $N$ states with generator $Q=\{Q_{ij}\}$, where $Q_{ij}\ge 0$ for $i\neq j$, $Q_{ii}=-\sum_{j\neq i}Q_{ij}$, and the master equation
\begin{equation}
\dot p_i(t)=\sum_{j=1}^N p_j(t)\,Q_{ji}.
\label{eq:master}
\end{equation}
Assume the uniform distribution $u_i=1/N$ is stationary:
\begin{equation}
uQ=0
\;\;\Longleftrightarrow\;\;
\sum_{i=1}^N Q_{ij}=0\quad(\forall j).
\label{eq:uniform-stationary}
\end{equation}
Define $\Omega_{\mathrm{eff}}(p)=\prod_i p_i^{-p_i}$, the Shannon entropy $S(p)=-\sum_i p_i\ln p_i=\ln\Omega_{\mathrm{eff}}(p)$, and
\begin{equation}
D_u(p)=\sum_{i=1}^N p_i\ln\frac{p_i}{1/N}
=\ln\!\frac{N}{\Omega_{\mathrm{eff}}(p)}.
\label{eq:Du-omega}
\end{equation}
We show that under \eqref{eq:master} with \eqref{eq:uniform-stationary},
\begin{equation}
\frac{d}{dt}D_u\big(p(t)\big)\;\le\;0,
\qquad
\frac{d}{dt}\,\Omega_{\mathrm{eff}}(t)\;\ge\;0.
\label{eq:Du-dec-Omega-inc}
\end{equation}

\textit{Proof by semigroup contractivity.} Let $P_t=e^{tQ}$ be the Markov semigroup. Then $P_t$ is stochastic and leaves $u$ invariant, $uP_t=u$. By the data-processing inequality for Kullback–Leibler divergence,
\begin{equation}
D_u\big(p(0)P_t\big)=D\!\big(p(0)P_t\ \Vert\ uP_t\big)\;\le\;D\!\big(p(0)\ \Vert\ u\big)\qquad(\forall t\ge 0).
\end{equation}
Setting $p(t)=p(0)P_t$ yields $D_u\big(p(t)\big)\le D_u\big(p(0)\big)$ for all $t$, hence $\tfrac{d}{dt}D_u\big(p(t)\big)\le 0$ almost everywhere.

From \eqref{eq:Du-omega} we have $D_u=\ln N - S$ with $S=\ln\Omega_{\mathrm{eff}}$, so
\begin{equation}
\dot D_u(t)= -\,\dot S(t) = -\,\frac{d}{dt}\,\ln\Omega_{\mathrm{eff}}(t).
\end{equation}
Therefore $\dot D_u\le 0$ implies
\begin{equation}
\frac{d}{dt}\,\ln\Omega_{\mathrm{eff}}(t)\;\ge\;0
\qquad\Longleftrightarrow\qquad
\frac{d}{dt}\,\Omega_{\mathrm{eff}}(t)=\Omega_{\mathrm{eff}}(t)\,\frac{d}{dt}\,\ln\Omega_{\mathrm{eff}}(t)\;\ge\;0,
\end{equation}
showing that the effective number of accessible states grows monotonically whenever the uniform distribution is stationary.

\textit{General reference law.} For a general stationary distribution $\pi$, the nonadiabatic entropy-production rate satisfies
\begin{equation}
\dot e_p^{\mathrm{na}}(t)
=\frac{1}{2}\sum_{i\neq j}\!\Big(p_i(t)Q_{ij}-p_j(t)Q_{ji}\Big)\,
   \ln\frac{p_i(t)\,\pi_j}{p_j(t)\,\pi_i}\;\ge\;0,
\end{equation}
and
\begin{equation}
\frac{d}{dt}\,D_\pi\big(p(t)\big)=-\dot e_p^{\mathrm{na}}(t)\;\le\;0.
\end{equation}
Specializing to $\pi=u$ recovers $\dot D_u(t)=-\dot e_p^{\mathrm{na}}(t)\le 0$ and hence $\dot\Omega_{\mathrm{eff}}(t)\ge 0$.

\section*{Acknowledgment}
I would like to thank  Mulu  Zebene  and Asfaw Taye for their
constant encouragement.

\end{document}